\documentclass{article}
\usepackage[a4paper, total={6.7in, 9.7 in}]{geometry}
\usepackage{cite}
\usepackage{url}
\usepackage{array}
\usepackage{amsmath,amssymb,amsfonts}

\usepackage{algorithm,algorithmic}
\usepackage{graphicx}
\graphicspath{ {./figures/} }
\usepackage{textcomp}
\usepackage{xcolor}

\date{}
\usepackage{authblk}

\begin{document}
\title{Exploring Blockchain Interoperability: Frameworks, Use Cases, and Future Challenges}
% \author{Stanly Wilson, Kwabena Adu-Duodu, Yinhao Li, Ellis Solaiman, Omer Rana, Rajiv Ranjan}
\author[1,3]{Stanly Wilson}
\author[1]{Kwabena Adu-Duodu}
\author[1]{Yinhao Li}
\author[1]{Ellis Solaiman}
\author[2]{Omer Rana}
\author[1]{Rajiv Ranjan}

\affil[1]{Newcastle University, UK}
\affil[2]{Cardiff University, UK}
\affil[3]{St Vincent Pallotti College of Engineering \& Technology, Nagpur, India}

% \author{
%     \IEEEauthorblockN{
%         Stanly Wilson\IEEEauthorrefmark{1}\IEEEauthorrefmark{3},
%         Kwabena Adu-Duodu\IEEEauthorrefmark{1},
%         Yinhao Li\IEEEauthorrefmark{1},
%         Ringo Sham\IEEEauthorrefmark{1},
%         Ellis Solaiman\IEEEauthorrefmark{1},        
%         Omer Rana\IEEEauthorrefmark{2},
%         Rajiv Ranjan\IEEEauthorrefmark{1}
%     }
%     \IEEEauthorblockA{
%         \IEEEauthorrefmark{1}Newcastle University, UK\\
%         \IEEEauthorrefmark{2}Cardiff University, UK\\
%         \IEEEauthorrefmark{3}St Vincent Pallotti College of Engineering \& Technology, Nagpur, India
%  %       
%     }
% }

% \address[add1]{Newcastle University, UK}
% \address[add2]{Cardiff University, UK}
% \address[add3]{St Vincent Pallotti College of Engineering \& Technology, Nagpur, India}
\maketitle
\begin{abstract}
% trust aspect
% data sharing problem
% discuss platforms
% Polkadot in detail
% case study
% future challenges
Trust between entities in any scenario without a trusted third party is very difficult, and trust is exactly what blockchain aims to bring into the digital world with its basic features. Many applications are moving to blockchain adoption, enabling users to work in a trustworthy manner. The early generations of blockchain have a problem; they cannot share information with other blockchains. As more and more entities move their applications to the blockchain, they generate large volumes of data, and as applications have become more complex, sharing information between different blockchains has become a necessity. This has led to the research and development of interoperable solutions allowing blockchains to connect together. This paper discusses a few blockchain platforms that provide interoperable solutions, emphasising their ability to connect heterogeneous blockchains. It also discusses a case study scenario to illustrate the importance and benefits of using interoperable solutions. We also present a few topics that need to be solved in the realm of interoperability.
\end{abstract}

\section{Introduction}
\par Blockchain has created new opportunities for the creation of applications that can foster trust between the involved parties. The initial growth of blockchain technology was largely restricted to cryptocurrencies. Later, several applications began to investigate the potential uses of blockchain for their operations; these applications primarily used blockchain for provenance and tracking. Prior to blockchain, systems attempted to externally enforce trust without built-in mechanisms. With the advent of blockchain, the element of trust (immutability) was introduced into the system, eliminating the requirement for external entities (third parties). Blockchain achieves trust using inherent mechanisms such as smart contracts, distributed ledgers, consensus algorithms, and cryptography. As blockchain became more widely used, several platforms appeared that offered features tailored to the needs of various organisations. However, no single blockchain platform can meet all the demands of the industry, leading to the emergence of numerous blockchain platforms, each tailored to meet a different set of needs. Each has its own advantages and disadvantages.
% Multi-layer Blockchain (a solution to the single-layer blockchain)
% The drawbacks of single-layer blockchains pushed the need for multilayer blockchains, and a few solutions have already come up to bridge the gap. 
\par The developments of blockchain solutions/applications often face the challenge of choosing between decentralisation, security and scalability; this is commonly referred to as the blockchain trilemma. The trade-off between these three is a real challenge, meaning that achieving an optimal balance between these three properties remains challenging in practice, often requiring trade-offs depending on application requirements. Most often, applications built using blockchain have stressed the aspect of decentralisation and neglected scalability. Due to this, applications are often less scalable than their standalone counterparts~\cite{blockchain_trilema}. Many consensus protocols and blockchain platforms have been developed to solve this blockchain trilemma, which remains mostly on layer-1 blockchains. 
% This raises the question of what are layer-0 and layer-1 blockchains. Bitcoin and Ethereum are two familiar names in the blockchain domain. They are layer-1 blockchains built on layer-0 for the crypto environment. 
Layer-0 blockchains are the base that provides the underlying infrastructure and function to create new chains and permit interoperability. It provides developers with the features required to create efficient applications. Layer-0 blockchains are general purpose, and application-specific blockchains can be built upon them. Layer-1 blockchains benefit from layer-0, and blockchains built on this can have specific consensus mechanisms and additional security features to ensure chain safety. However, the second layer does not have inherent mechanisms for cross-chain communication. It also has issues with scaling applications and provides low throughput. To overcome the drawbacks of layer-1 blockchains, there are layer-2 blockchains to support better throughput and scalability~\cite{layer0}. 

% \textbf{Motivation}
\par Many organisations in the initial states explored blockchain individually and now need ways to share information with their industry partners who may have applications built on other blockchain platforms~\cite{interoperability_2}. Consider the scenario of a supply chain where multiple organisations/entities participate and exchange information. Let these entities be a food-grain industry, a logistic company to transport the products, a grocery chain retail company, and a financial payment service. For the efficient operation of such a supply chain, information must flow between entities despite their underlying systems. Since no single blockchain platform can support all application requirements, different entities often adopt platforms tailored to their specific needs. In this scenario, with existing traditional blockchains, information sharing becomes very challenging since each blockchain may have different underlying security features and consensus protocols. Consider another scenario in healthcare that has several entities, such as hospitals/health clinics, health monitoring services, pharmacy chains, payment channels, and insurance/claims services. The hospitals/clinics have the health details of the patients, and they may use health monitoring devices to track the health metrics of their patients. They have to share the medication details of the patients with pharmacy chains, where the patients can collect their prescriptions and pay through some payment channels. In situations with some claims or insurance, the information must also flow to these entities. Information sharing becomes challenging if these entities develop their solutions on separate blockchains. These scenarios point to the need for interoperability among blockchains. The drawbacks of single-layer blockchains have pushed the need for multilayer blockchains, and a few solutions have already come up to bridge the gap. 

{Several recent studies have explored blockchain interoperability from different perspectives. For example, Belchior et al. \cite{interoperability_2} provide a comprehensive survey of interoperability protocols and classifications, while works such as Ren et al. \cite{general_ch_1} and Kotey et al. \cite{std_2} examine architectural designs, communication models, and challenges in heterogeneous blockchain environments. Unlike these existing surveys, which primarily focus on interoperability protocols, classifications, and taxonomies, this paper adopts a platform-centric perspective.} Rather than providing an exhaustive survey, we focus on how widely used blockchain ecosystems implement interoperability mechanisms in practice. In addition, we complement this analysis with a conceptual multi-entity use case to illustrate how interoperability is realised in real-world scenarios. This combined perspective provides practical insights into the design and application of interoperable blockchain systems. The contributions of this paper are as follows. 

% \textbf{Contributions} 
\begin{itemize}
    % \item We make a general study of the major platforms that provide interoperability between heterogeneous blockchain platforms.
    % \item We propose a supply chain-based case study to demonstrate how interoperable blockchains can enhance a specific use case.
    % \item We list key challenges that need to be resolved and require further research. 
    \item {A platform-oriented analysis of widely used blockchain ecosystems that support interoperability, highlighting the architectural mechanisms used to enable cross-chain communication.}
    \item {A conceptual supply-chain case study demonstrating how heterogeneous blockchain systems can interact using interoperable architectures.}
    \item {A discussion of open research challenges related to interoperability in heterogeneous multi-chain environments.}
\end{itemize}

\par Unlike prior survey-based studies, this work emphasises a platform-oriented analysis combined with a practical interoperability scenario, providing a clearer understanding of how interoperability mechanisms are applied in real-world settings.

\par The structure of the rest of the paper is as follows. Section~\ref{sec: interoperability} deals with the interoperable solutions proposed for homogeneous and heterogeneous blockchain platforms. Section~\ref{sec: scenario} discusses a case study on a supply chain scenario from the perspective of an interoperable blockchain. Section~\ref{sec: future_challenges} discusses some challenges that require further attention. Finally, Section ~\ref{sec: conclusion} concludes the paper.
% Literature review
% 1. A Survey on Blockchain Interoperability: Past, Present, and Future Trends
% 2. Towards Interoperable Blockchains: A Survey on the Role of Smart Contracts in Blockchain Interoperability
% 3. Towards Blockchain Interoperability

\section{Interoperability in Blockchain} \label{sec: interoperability}
% \textbf{Definition of interoperability ?}
% •	Communicate (transact and change state) between multiple homogeneous or heterogeneous blockchains having their unique ledgers 
% •	Verifiable and traceable 

\par Interoperability can be defined as the ability of blockchain systems to exchange data, assets, and state information across heterogeneous or homogeneous networks while ensuring verifiability and consistency of transactions across participating chains. While discussing the interoperability of blockchain, it is assumed that there are source and destination blockchains. In the initial stages, interoperability aimed to provide features for cryptocurrency systems. The paper \cite{ACM_survey_1} discusses two types of protocols that enable blockchain interoperability. The first is the cross-chain communication protocol (CCCP), which refers to communication between two homogeneous blockchains with similar features and underlying protocols. The second is the cross-blockchain communication protocol (CBCP), which refers to communication between homogeneous and heterogeneous blockchains or between two heterogeneous blockchains. Most interoperability solutions fall under these two categories. 
\subsection{Cross-Chain Communication Protocol}
\par One of the commonly used CCCP is sidechains. It is a mechanism that enables blockchains to communicate with each other and provides scalability features. In this case, one blockchain considers the other as an extension of itself. The mainchain contains the ledger assets, and the sidechain can access them. Sidechains improve the performance of the mainchain by offloading and processing transactions separately and returning the results to the mainchain. The essential components of the sidechain are CCCP and consensus protocols. 

The most common way in which sidechains interact with mainchains is called a two-way peg. If a client on the mainchain desires to transfer tokens to another party, it must first send those tokens to specific nodes on the mainchain. These nodes then lock the tokens within the mainchain and produce equivalent tokens on the sidechains \cite{interoperability_3_IEEE_Access}. The client may then use the newly created sidechain tokens to conduct transactions with the target blockchains. Each time the client utilises sidechain tokens, the locking nodes eliminate the corresponding tokens that are secured on the mainchain. This procedure is known as a two-way peg, with three prevalent implementations: simplified payment verification (SPV), central two-way peg, and federated two-way peg.

SPV clients are lightweight clients that do not store the full blockchain state but verify transactions using block headers and Merkle proofs. The central two-way peg uses a trusted central entity that acts as a communication channel. Although it performs well, it permits the centralisation of the process. The federated two-way peg overcomes centralisation by having a group of nodes to lock and unlock the tokens \cite{ACM_survey_1}. 
\par However, sidechains are layer-2 solutions built over layer-1 (blockchains) to support decentralised applications (dApps). However, sidechains have a few drawbacks. The basic assumption of a sidechain is that the mainchain is secure. Insecurity in the mainchain would invalidate the transactions performed on the sidechains. Secondly, decentralisation and performance are inversely proportional. If the system is more decentralised, it will affect its performance and scalability. Sidechains are often centralised to achieve better performance, which would lead to a single point of failure if an attacker gains control of the system. Conversely, if sidechains have a more robust security mechanism, it will lead to slower transactions and affect performance. Lastly, designing complex applications using sidechains is challenging as they do not provide features to specify conditions for the two-way peg mechanism to suit the requirements \cite{ACM_survey_1}. 
\par Similar to sidechains, other solutions use CCCP to enable interoperability, such as notary schemes and hashed time-lock contracts (HTLC). Notary schemes (exchanges) monitor other chains and enable chains to transact between other chains. HTLC achieves interoperability by providing time locks on contracts that are valid for only a specified time. Similar to sidechains, these solutions also have similar drawbacks \cite{interoperability_3_IEEE_Access,layer_2_survey}. 
% \\
\subsection{Cross-Blockchain Communication Protocol}
\par A few platforms support cross-blockchain communication that enable asset transfer between blockchains. Relevant literature was identified through searches in major academic databases, including IEEE Xplore, ACM Digital Library, Scopus, and Google Scholar, using keywords such as blockchain interoperability, cross-chain communication, multi-chain architectures, and heterogeneous blockchains. The blockchain platforms considered in this study were selected to represent diverse architectural approaches to interoperability, including relay chain models, hub-and-zone architectures, bridge-based communications, and subnet-based systems. Additionally, the selection was limited to platforms with well-documented interoperability mechanisms and significant ecosystem adoption, ensuring that the analysis reflects representative and practically relevant blockchain interoperability frameworks. In addition to peer-reviewed publications, technical whitepapers and official platform documentation were also consulted, as many blockchain interoperability frameworks are primarily described in developer documentation rather than academic literature. Accordingly, the platform descriptions in this section combine evidence from published studies and official technical documentation. Where interoperability mechanisms have been validated in prior studies, we describe them as implemented approaches; where claims are primarily based on platform documentation or whitepapers, they should be interpreted as documented architectural capabilities that may still be evolving in practice. The selected platforms are not intended to be exhaustive but are chosen to represent distinct and widely adopted architectural approaches to interoperability, allowing comparative insights across different design paradigms.

This section explores such platforms based on their capacity for interoperability, which includes interoperable entities and interoperable enablers. Interoperable enablers refer to the mechanisms, technologies, or protocols that enable communication, interaction, and data transfer between different blockchain systems. These are the tools or frameworks that make interoperability possible. An interoperable entity refers to the specific elements or units within or connected to a blockchain ecosystem that participates in or enables cross-chain communication. The blockchain platform comparison focuses primarily on architectural and interoperability-related dimensions, as these aspects directly influence the ability of blockchain platforms to support cross-chain communication.

% Ethereum
\subsubsection{Ethereum}
Ethereum \cite{layer_2_survey, ethereum_2.0} is one of the most widely used and flexible blockchain platforms that support developments. In 2022, a merger was conducted that enabled it to be more energy-efficient and scalable. It achieves interoperability and scalability through layer-2 solutions, such as rollups and bridges \cite{rollups}. Rollups enable off-chain computations and data storage while periodically submitting proofs to the Ethereum mainnet, thereby maintaining the security and decentralisation of the Ethereum network. Two types of rollups are commonly used: optimistic rollups, which assume validity and rely on fraud proofs, and zero-knowledge rollups (zkRollups), which use cryptographic proofs to ensure their validity. Ethereum uses bridges to facilitate communication with other blockchains and enable asset transfers. 

% hyperledger
\subsubsection{Hyperledger}
Hyperledger \cite{hyperledger_whitepaper,hyperledger} is an umbrella project of open-source blockchains and related tools initiated by the Linux Foundation. It provides a modular architecture that allows for pluggable consensus mechanisms, such as Raft and Byzantine Fault Tolerant (BFT) protocols, tailored to specific enterprise needs. To address interoperability challenges, Hyperledger offers tools such as Hyperledger Cacti \cite{hyperledger_cacti}, which is a pluggable interoperability framework designed to link networks built on heterogeneous blockchains. Hyperledger Cacti allows blockchains such as Hyperledger Fabric, Corda, and Ethereum to interoperate without requiring centralised intermediaries. Channels provide isolated environments for transaction processing and data sharing, whereas networks enable interoperability across heterogeneous blockchains.

% Cosmos
\subsubsection{Cosmos}
\par Cosmos \cite{cosmos} aims to create `internet of blockchains' that can communicate with each other. It consists of three layers: Application layer, Consensus layer and Networking layer. These three layers exchange and transfer information using the three underlying components of Cosmos: Tendermint, Cosmos SDK, and IBC. Tendermint BFT is a Byzantine fault tolerance (BFT) based consensus protocol that can achieve fast consensus in the creation of blocks. Tendermint BFT is application-agnostic and can access the application layer through the application blockchain interface (ABCI). ABCI can be implemented using any programming language enabling developers to use the programming language of their choice. Cosmos SDK is a framework that simplifies the creation of blockchains for various use cases. Cosmos achieves interoperability using the inter-blockchain protocol (IBC), which can communicate between Tendermint and non-Tendermint-based blockchains. Cosmos can transfer information between two independent blockchains,  referred to as zones, using connectors called hubs. It can connect blockchains with fast finality using hubs, while using pegs (similar to sidechains) to communicate blockchains with probabilistic finality (for example Bitcoin and Ethereum).

% Ark
\subsubsection{Ark}
\par Ark \cite{ark, ark_webpage}  aims to simplify the creation of blockchains by providing easily customisable blockchains and reducing deployment time and programming complexity. Ark was developed to overcome the blockchain trilemma and to support interoperability and sustainability. Ark tries to achieve these objectives by enabling interoperability and using SmartBridge to connect other blockchains, such as Bitcoin and Ethereum. SmartBridge supports two types of communication: it can connect Ark with blockchains built on Ark's core technology (common consensus protocol) and it can connect blockchains outside Ark (different consensus protocols). The former is a protocol-specific SmartBridge, whereas the latter is a protocol-agnostic SmartBridge.  Blockchains built on Ark core are called bridgechains, which enable entities to share information. A SmartBridge mechanism called Ark Contract Execution Services (ACES) enables communication between Ark and other blockchains. It uses a special data section called \textit{vendor field} and a set of \textit{encoded listeners} to process this data. Ark’s mainnet remains at the centre to which SmartBridge connects with other blockchains. SmartBridge can offload complex operations from the mainnet and push the results back to Ark, similar to the side-chain process. Ark provides an easily customisable blockchain SDK that can clone Ark’s mainnet and support developers with multiple programming language options. Ark uses a modified version of the Delegated Proof of Stake (DPoS) as the consensus protocol.

% Harmony
\subsubsection{Harmony}
Harmony \cite{Harmony_whitepaper} is a fast and energy-efficient blockchain platform that focuses on scalability and interoperability. It uses a unique consensus protocol called Effective Proof of Stake (EPoS), which reduces centralisation risks while enabling fast block creation (2 seconds with finality) and low transaction fees. Harmony incorporates sharding to process transactions in parallel, thereby improving scalability. It achieves interoperability through the LayerZero Bridge, which connects Harmony with blockchains such as Ethereum, Binance Smart Chain, and Bitcoin. LayerZero Bridge enables the transfer of assets, NFTs, and data between these networks, promoting seamless interaction across blockchains. The platform’s interoperability is further enhanced through its ability to connect shards within the Harmony ecosystem and with external blockchains \cite{Harmony_docs_web}. 

% Avalanche
\subsubsection{Avalanche}
\par Avalanche \cite{avalanche_1} blockchain aims to solve the issues related to scalability and tries to improve performance. It provides intense competition to Ethereum and supports solidity-based smart contracts out of the box. Most blockchains take a few seconds or minutes to create blocks, whereas Avalanche does so in a second. Avalanche overcomes the scalability issue and enables interoperability using subnets or subnetworks (a set of validators) that create and maintain independent and self-governed blockchains while sharing the security features of the Avalanche platform. Avalanche does not impose any limit on the number of subnets in its platform. The Avalanche platform has three constituting chains. First is the contract chain (c-chain), which is responsible for creating smart contracts. It uses an instance of the Ethereum Virtual Machine (EVM) and supports Ethereum-based applications. The second is the platform chain (p-chain), which tracks the subnets. The exchange chain (x-chain) is the third chain that creates and transfers assets and tokens in the chain. Avalanche employs two consensus protocols to secure these chains. The Avalanche consensus protocol secures the x-chain while the others use the snowman consensus protocol, a modified version of the Avalanche consensus protocol \cite{avalanche_2}. {Avalanche’s popularity and performance features have been widely recognised; however, reports in the media have raised questions regarding aspects of its decentralised nature} \cite{avalanche_issue}.

% near protocol
\subsubsection{Near}
Near \cite{near_protocol_whitepaper} is a blockchain platform that combines sharding and advanced consensus mechanisms to achieve both interoperability and scalability. It employs a sharding model called Nightshade, which splits the blockchain into multiple shards, each processing a subset of transactions, thereby enabling parallel processing and high throughput. The Near protocol achieves cross-chain interoperability through the Rainbow Bridge, which connects Near with Ethereum. Rainbow Bridge allows users to seamlessly transfer assets and data between the two networks. It uses smart contracts on both chains to validate and relay information, thereby ensuring security and decentralisation. Additionally, Near uses its Aurora engine, an EVM-compatible environment, to execute Ethereum-based smart contracts on the Near blockchain \cite{near_docs}.

% Solana
\subsubsection{Solana}
Solana \cite{solana_whitepaper} is a high-performance blockchain designed for scalability and low latency. It uses a novel consensus mechanism called Proof of History (PoH) combined with Tower Byzantine Fault Tolerance (BFT) to achieve fast transaction finality and high throughput \cite{solana_consensus}. Interoperability in Solana is facilitated by the Wormhole Bridge, a decentralised cross-chain messaging protocol. Wormholes enable seamless transfer of assets and information between Solana and other blockchains. The bridge uses a network of validators to verify cross-chain messages and secure asset transfers without the need for a central authority. This interoperability enabler allows Solana-based decentralised applications to interact with assets and services on other blockchains, broadening its ecosystem.

% Cardano
\subsubsection{Cardano}
\par Cardano \cite{cardano} is a blockchain platform that evolved to solve the scalability and performance bottlenecks of Ethereum. It uses a modified version of Proof of Stake (PoS) called 'Ouroboros' to reach consensus in the network. Cardano aims to improve the performance of blockchains by having more transactions at lower costs and attempts to achieve this by separating transactions from the computation. Cardano has two operational layers: the Cardano Settlement Layer (CSL) and the Cardano Computation Layer (CCL). The CSL has information regarding accounts and balances, and manages the assets in the network. CCL deals with the execution of smart contracts and other computational activities. For the development process, Cardano proposes two languages: \textit{Plutus} and \textit{Marlowe}. \textit{Plutus} is Cardano’s smart contract language and it is based on \textit{Haskell}, a functional programming language. \textit{Marlowe} is a domain-specific language to support decentralised finances. Cardano uses node-to-node inter-process communication (IPC) to share transaction details with other entities (sidechains) which are connected to its mainnet. 

% Algorand
\subsubsection{Algorand}
Algorand \cite{algorand_whitepaper} is a highly secure and scalable blockchain platform that employs a unique Pure Proof of Stake (PPoS) consensus mechanism, where validators are randomly selected based on the weight of their stake, ensuring fast and secure transaction finality while maintaining decentralisation \cite{algorand_consensus}. Algorand achieves interoperability through state proofs and bridges, which enable secure and efficient communication with external blockchains. State proofs are lightweight cryptographic proofs that allow Algorand to verify transactions and state changes from other blockchains without requiring a trusted intermediary. Algorand’s layer-1 architecture natively supports interoperability, making it easier to directly integrate cross-chain functionality within its ecosystem. Additionally, Algorand bridges facilitate the transfer of assets and data between Algorand and networks such as Ethereum \cite{algorand_interoperability}.

% Polkadot
\subsubsection{Polkadot} \label{sec: polkadot}
\par Polkadot \cite{Polkadot_wp} is a “scalable heterogeneous multichain” that provides “globally coherent dynamic data structures” called parachains. As the name suggests, parachains are “parallelised chains” that participate in the Polkadot network and are independent entities. 
% Polkadot has solved two issues with blockchain development for quite some time: interoperability and blockchain trilemma (layer-1 concern) because Polkadot is a layer-0 solution. 
The relay chain is the foundation of the Polkadot network, allowing independent blockchains to communicate for true decentralisation. It processes transactions from all chains simultaneously and provides a shared security model through its consensus mechanism. Parachains are customisable blockchains for specific applications, gaining interoperability and security through the Cross-Consensus Message Passing (XCMP) protocol via a relay chain. Bridges connect blockchain networks for data transfers. Not all networks need to be part of Polkadot; bridges enable asset exchanges and interoperability with external networks. Validator nodes uphold the relay chain by creating and verifying blocks for each parachain. They stake personal funds under the NPoS (Nominated Proof of Stake) algorithm to ensure proper conduct, earning points for valid parachain block confirmations. The relay chain provides shared security for connected blockchains while enabling decentralised cross-chain communication via XCMP. It uses BABE and GRANDPA \cite{polkadot_consensus} for consensus. Polkadot ensures interoperability with external networks, such as Bitcoin and Ethereum, through bridges that transfer data and assets \cite{polkadot_2022}.

Table~\ref{tab: interoperability_table} summarises the interoperable blockchains discussed in this section along with their key features and characteristics. This comparison highlights the differences in architectural design, consensus mechanisms, and interoperability enablers used to support cross-chain communication. These differences reflect the diverse strategies adopted by blockchain ecosystems to achieve interoperability while balancing scalability, security, and system flexibility.
% \begin{sidewaystable}
\begin{table}[h!] 
\begin{center}
\centering
\caption{Comparative Overview of Interoperable Blockchain Platforms and their Interoperability Mechanisms}
\label{tab: interoperability_table}
\begin{tabular}{ | m{2.5cm} | m{3cm}| m{3cm} |  m{3cm} | m{2.5cm} |}
 \hline
 \textbf{Blockchain} & \textbf{Consensus Protocol} & \textbf{Main chain} & \textbf{Interoperability enabler} & \textbf{Interoperable entity} \\  [0.5ex] 
 \hline
 Ethereum & Proof of Stake (PoS) & Ethereum Mainnet & Rollups, Bridges & Layer 2 Chains, Bridges \\ 
 \hline
 Hyperledger & Pluggable Consensus & Public and Private Chains & Hyperledger Cacti & Channels and Networks \\ 
 \hline
 Cosmos & Tendermint BFT & Cosmos Hub & IBC & Zones and Hubs \\ 
 \hline
 Ark & DPoS & Ark Mainnet & SmartBridge & Bridgechains \\
 \hline
 Harmony & Effective Proof of Stake (EPoS) & Harmony Mainnet & LayerZero Bridge & Shards \\
 \hline
 Avalanche & Avalanche and Snowman & Primary Network & p-chain & Subnetworks/ Subnets \\
 \hline
 Near & Sharded Proof of stake (Nightshade) & NEAR Mainnet & Rainbow Bridge & Shards \\ 
 \hline
 Solana & Proof of History (PoH) and Tower BFT & Solana Mainnet &Wormhole Bridge & Cross-Chain Tokens \\ 
 \hline
 Cardano & Ouroboros & Cardano Mainnet & Node-to-node IPC & Sidechains \\
\hline
 Algorand & Pure Proof of Stake (PPoS) & Algorand Mainnet &State Proofs and Bridges & Layer 1 Architecture \\
 \hline
 Polkadot & BABE and GRANDPA & Relay chain & XCMP & Parachains \\ [1ex] 
 \hline
\end{tabular}
\end{center}
\end{table}

\section{Use Case Scenario: Supply Chain} \label{sec: scenario}
% Why supplychain information sharing is needed?
% Why the use of blockchain in supplychain?
% Explain the roles of the entities
% Explain what functions the model can achieve.
% Explain the general architecture (explain the things considered and not like collators, validators etc.)
% Explain the smart-contract logic if possible
Understanding the practical benefits of interoperability solutions requires examining real-world scenarios where such platforms provide transformative value. This section presents a supply chain use case to illustrate how leveraging the capabilities of interoperable blockchains can overcome traditional bottlenecks and unlock new possibilities in decentralised ecosystems. The purpose of this case study is to illustrate how interoperable blockchain architectures can support interactions among multiple heterogeneous systems. This case study is conceptual in nature and is intended to demonstrate architectural feasibility rather than provide an empirical performance evaluation. Although no prototype implementation or quantitative benchmarking was conducted, the scenario highlights how interoperability mechanisms can coordinate multi-entity interactions in complex environments. The survey \cite{ACM_survey_1} indicates that Cosmos and Ark are typically limited to linking a small number of heterogeneous blockchains, whereas Polkadot supports multiple heterogeneous chains through its relay chain architecture. This makes Polkadot suitable for illustrating complex multi-stakeholder scenarios, such as supply chain ecosystems \cite{usecase_1}. Having discussed Polkadot and its architecture in Section ~\ref{sec: polkadot}, we consider a concrete scenario that illustrates the benefit of using parachains over standalone blockchains and detail the various possibilities Polkadot provides to make interoperability possible. 

In recent years, the use of blockchains in supply chains (SC) has attracted attention from both industry and academia. Traditional SCs use centralised management systems to ensure data integrity and security, while risking system corruption and data tampering. Although SCs use various traditional methods, such as storing data digitally, sharing information, and tracking products, there have been many events \cite{grain_sc_1} in the past in which they faced challenges and threats owing to data tampering, information loss, and false SC entities. Such incidents demonstrate the weaknesses of traditional SC methods, necessitating better tracking of products and materials in the SC. Another area in which SCs face challenges is the cumbersome process required to trace the origin or journey of products in the SC ecosystem \cite{sc_1}. Much research has been conducted in the last few years on blockchain and SC integrations. The paper \cite{sc_2} identified that data sharing requires mutual trust between the entities/parties involved. The SC ecosystem requires more auditability and product traceability to be more productive and to improve overall performance. Their analysis proved that sharing correct data on a blockchain can also increase economic benefits. Undoubtedly, blockchain can improve the SC ecosystem with general auditability; however, interoperability and scalability have always been issues with layer-1 blockchain solutions.

\par
\begin{figure}[!ht]
\centerline{\includegraphics[scale=0.6]{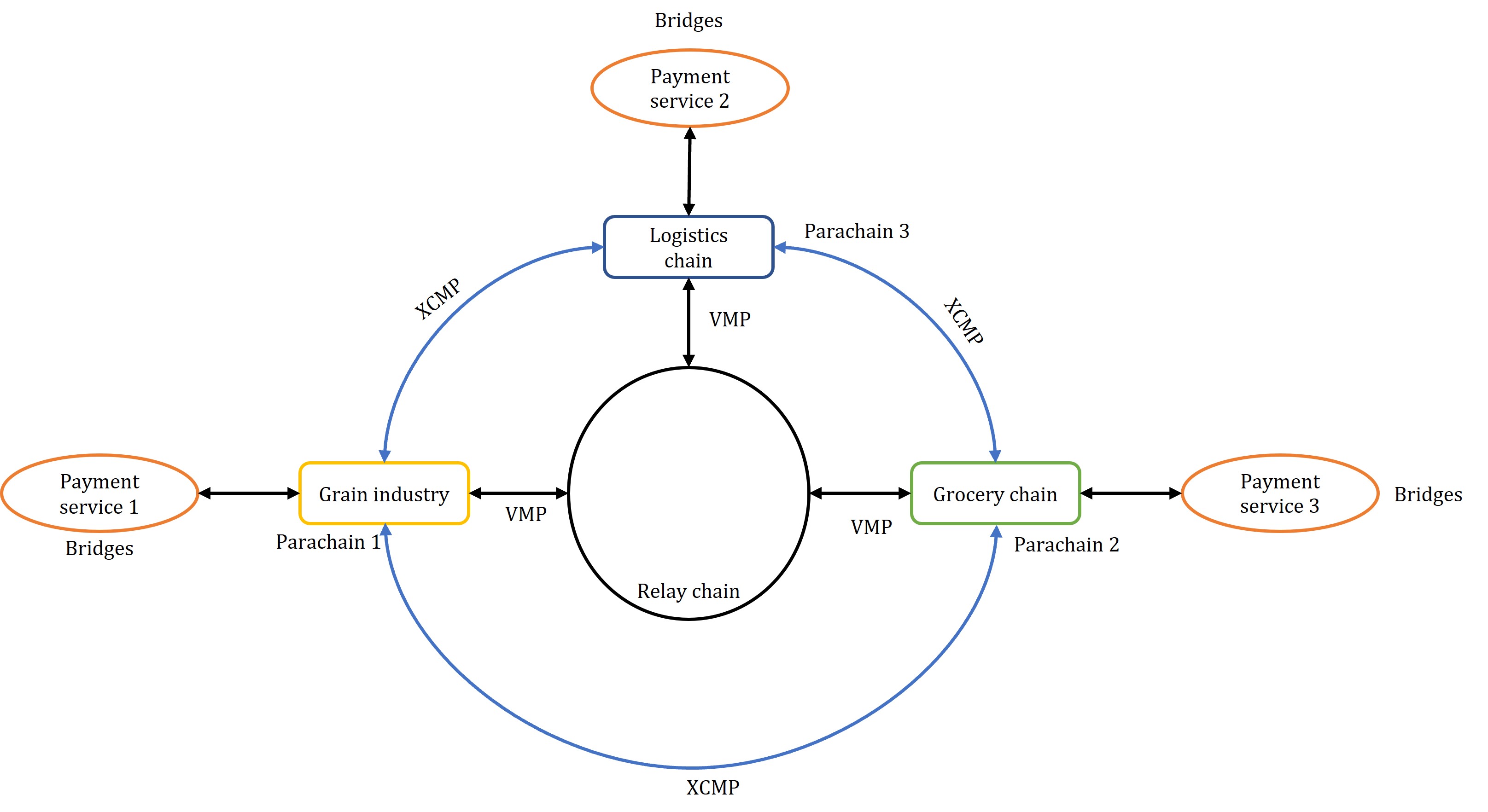}}
\caption{Architecture of the Interoperable Blockchain-based Supply Chain Scenario}
\label{architecture}
\end{figure}

 Applying the parachains in the SC setting, we try to demonstrate how Polkadot can overcome the interoperability and scalability issues. In our case study, we consider four entities. 1) Food/grain industry, 2) logistics company for which the first entity is one of its many clients, 3) grocery retail chain who are one of the many consumers of the first entity and 4) payment services through which the financial transactions can take place. We assume that the three entities (food, logistics and grocery) are heterogeneous blockchains hosted as three parachains and financial services are bridges. Figure \ref{architecture} illustrates the architecture of our scenario. 

\par In the given scenario, the entities can perform the following functions: a) The entities can transact/ share information and the proof of which can be placed in the relay chain, b) the entities can perform financial transactions through their choice of available payment services (bridges), and the receiver can receive it in their end, c) the proof of all the transactions between the parachains will be available in the relay chain which stands as the guarantee of transaction and each entity can individually verify the provenance and d) at a given point of time, the entities can track the status of the product/service. 

\par The parachains communicate with each other using XCMP. This communication involves several aspects. First, the sender parachain gossips the message to other nodes in the relay chain so that all nodes are aware of the transaction. Secondly, the collators of the receiver parachain receive the message, and the validators of the receiver parachain validate the message. Thirdly the validators create the validated message as a block and push it to the relay chain using vertical message passing (VMP). 

\par
\begin{figure}[!ht]
\centerline{\includegraphics[scale=0.6]{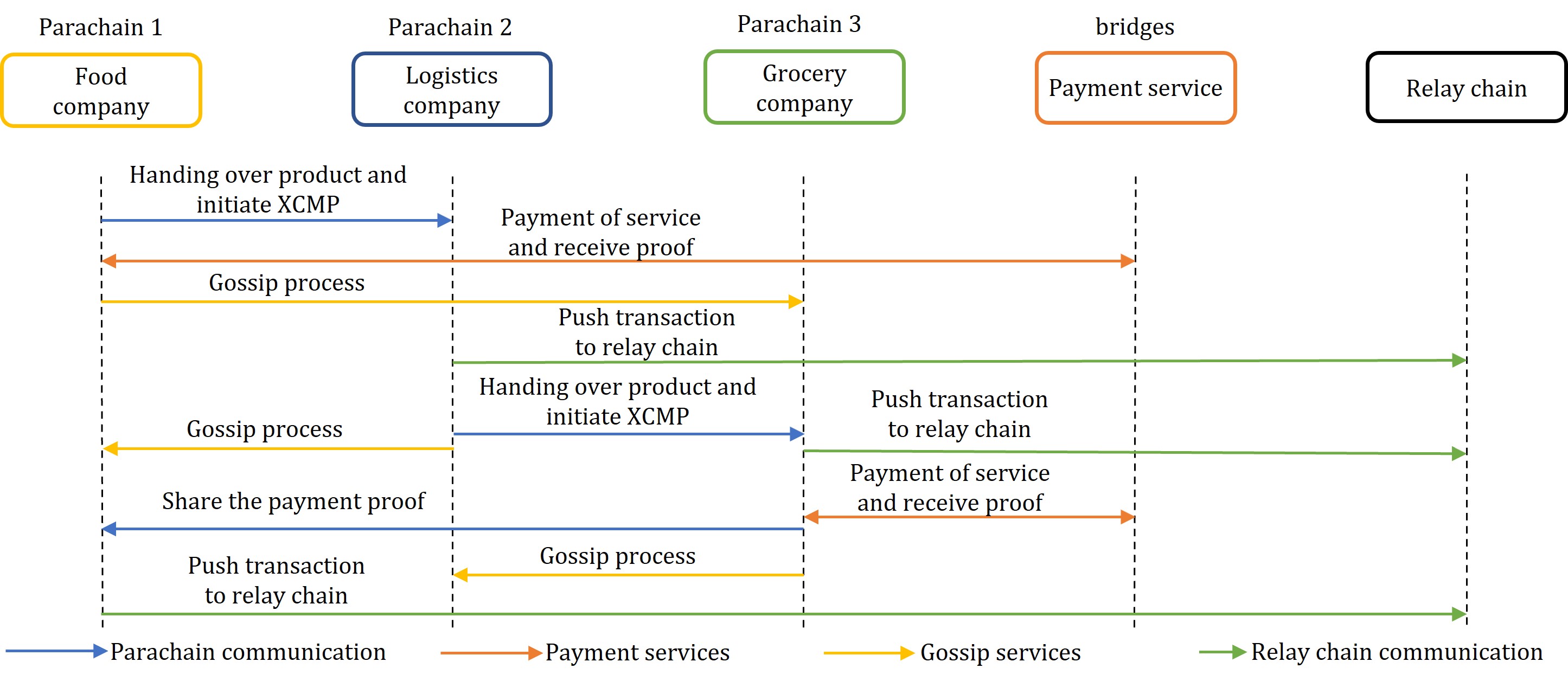}}
\caption{Sequence Diagram Illustrating Cross-chain Interactions between Supply Chain Entities (simplified interaction flow)}
\label{sequence_d}
\end{figure}

 The data flow starts with the food company (parachain 1) sending their products. The sequential flow of this process is as follows: The food company (parachain 1) prepare the product consignment and hand it over to the logistics company (parachain 2). The food company makes the payment for the services of the logistics company through the payment service and obtains the transaction proof. Parachain 1, using XCMP, communicates to parachain 2 that the consignment has been handed over and provides proof of payment. This XCMP communication triggers the gossip process and informs all nodes in the relay chain about the communication between parachain 1 and parachain 2. The collator of parachain 2 receives the message, and its validators would validate the message. The validators push the message to the relay chain using the VMP if they accept the message. The logistics company moves the consignment to the grocery chain (parachain 3) warehouse. The logistics company then communicates with the grocery chain regarding delivery. This triggers the gossip process and shares the transaction details with other nodes in the relay chain. Parachain 3 then validates the received product, and validators confirm the message from parachain 2. If the validators accept the transaction, the grocery company gains ownership of the products, and the transaction details are pushed to the relay chain as blocks. The grocery company makes the payment for the product from the food company through the payment service, receives the payment confirmation, and communicates to parachain 1, which triggers the gossip process. Parachain 1 receives the message, validates it, and pushes the transaction as a block in the relay chain. The parachains can access all the relay chain transactions enabling them to know the status of their transaction. Figure \ref{sequence_d} shows the sequence diagram of the process.

% \begin{enumerate}
%  \item The food company prepare the product consignment and hand it over to the logistics company. 
%  \item Make the payment through the payment service and obtain the transaction proof. 
% \item Using XCMP, communicate to the logistic company (parachain 2) that the consignment has been handed over and provide proof of payment. 
% \item The sender parachain (in this case, parachain 1) gossips the information to all the nodes.
% \item The collator of parachain 2 (logistics) receives the message, and its validators would validate the message. The validators are pushed to the relay chain using VMP if they accept the message. 
% \item The logistic company moves the consignment to the grocery chain warehouse. The logistic company then will communicate to the grocery chain (parachain 3) regarding the delivery. 
% \item Parachain 2 will perform step 4, and all the nodes will have this information. 
% \item Parachain 3 then validates the received product, and validators confirm the message from parachain 2. They pushed the validated information to the relay chain as blocks. 
% \item The grocery company makes the payment to the food company through the payment service, receives the confirmation of payment 
% \item Grocery company share with the food industry. 
% \item Gossip process of step 4 
% \item Parachain 1 receives the message, validates the same and pushes the transaction as a block in the raply chain
% \end{enumerate}

% figure configuration 

% 1. Interoperability
% 2. Freedom of choice
% 3. Proof of transaction on the relay chain
% 4. Underlying security mechanism
\par The above scenario of SC has the following advantages over the same on a single-layer blockchain. First of all, it provides interoperability. The entities hosted on different blockchains can communicate with each other. The blockchain platform on which the entities host their blockchains does not matter. It provides more freedom for clients to use a blockchain platform of their choice to build their solutions. Secondly, the relay chain stands as proof for all the parachain transactions that bring the aspect of trust while dealing with multiple entities and situations of conflict. All nodes have access to relay chain data that can be used to trace and verify transactions. Lastly, the Polkadot ecosystem has inbuilt security mechanisms that ensure the atomicity and validity of transactions.

A few studies \cite{e-science, market_place} have demonstrated the feasibility of similar interoperability architectures. They proposed a blockchain-enabled framework for the circularity and sustainability of materials in the built environment, in which the Polkadot ecosystem and its parachain architecture were utilised to enable interoperability between heterogeneous blockchain systems. The architectural principles used in that framework align with the interoperability approach illustrated in the supply chain scenario presented in this paper.

\section{Future Challenges} \label{sec: future_challenges}
The case study demonstrated the potential of parachains in providing interoperability between blockchains, which opens new avenues for implementing complex business logic on the blockchain with the freedom to choose any blockchain platform for development. Designing interoperable blockchain systems involves several important architectural trade-offs that must be considered. For example, mechanisms that prioritise strong security guarantees may introduce additional communication overhead, thereby affecting system efficiency. Similarly, increasing decentralisation can improve trust and resilience but may reduce performance owing to the coordination required across multiple networks. Interoperable systems must also balance openness and transparency with the need to protect sensitive data, particularly in enterprise or supply chain applications, where confidentiality is important \cite{general_ch_2}.

Although the interoperability platforms and case study discussed above enable cross-chain communication through different architectural approaches, each model presents certain trade-offs. Relay chain architectures such as Polkadot provide strong coordination and shared security but introduce additional architectural complexity. Hub-based models, such as Cosmos, enable the flexible interconnection of blockchains but rely on protocol compatibility. Bridge-based solutions offer interoperability between otherwise independent networks but may introduce additional security concerns. These differences highlight the trade-offs that blockchain ecosystems must consider when designing interoperable architectures \cite{general_ch_1}.

Each of the following challenges is discussed in terms of its impact on interoperability, along with references to emerging approaches that partially address these issues, highlighting areas where further research is required.

\subsection{Data Management}
% Blockchain differs from traditional databases in the case of storage
% What blockchain does not have in terms of storage
% Why does blockchain need storage?
% Why storing complex data is a concern and its query
The initial implementations of blockchain were mainly on crypto-based applications, which had fewer pieces of information to store on the blockchain. It mostly contained sender and receiver information and transferred amounts. Owing to its design, the blockchain data structure inherently deals with sender and receiver information and only needs to store the transaction amount. As the use cases and applications increased, storing data on the blockchain became expensive, leading to different storage options, such as on-chain (storing on the blockchain) and off-chain (storing outside the blockchain). Blockchain platforms are often designed for specific applications, resulting in isolated data environments that limit seamless data exchanges across different blockchain networks. Although cross-chain frameworks such as Polkadot and Cosmos enable communication between heterogeneous chains, managing and synchronising data across multiple blockchains introduces challenges related to data consistency, security, and reliable cross-chain transaction handling \cite{data_cha_1}. Developments in decentralised storage offer a solution to data storage issues on blockchain, and the most popular one in this category is the Inter-Planetary File System (IPFS) \cite{ipfs_2}. It supports storing information that can be easily accessed from anywhere, and it overcomes the single point of failure with its decentralised feature. Blockchain platforms, such as Polkadot, offer integration with IPFS to store the files. As the name suggests, IPFS is a file system; applications may need to store not only files but also various transaction data, product details, and other related records, which may require a database \cite{ipfs}. Emerging decentralised database systems, such as BigchainDB \cite{Bigchain}, show promise but require further development to seamlessly integrate with interoperable blockchains.

\subsection{Query Aggregation and Indexing}
% Why do we need query aggregation?
% Why does indexing help to query?
Following the data management challenge, there is another challenge in query aggregation and data indexing that must be addressed. Even if all data are stored on the blockchain (not advised), querying data depending on their attributes would cause a problem. Blockchain provides data provenance and auditability features using transaction details. They cannot perform a query on the stored data and return the results. From our case study, consider a situation in which parachain 1 needs to obtain all data details communicated with parachain 3. The present blockchain architecture does not support complex query features \cite{query2022,query_icbc}. 
Additionally, cross-chain queries pose unique challenges. The diversity of data structures and formats across interoperable blockchains complicate query standardisation. Solutions such as the Graph Protocol \cite{graph_protocol}, which indexes blockchain data, offer potential but are limited in their ability to handle heterogeneous data models. To query effectively, systems must also know what data exist and where they reside, highlighting the need for advanced indexing mechanisms that support federated queries across multiple blockchains \cite{query_ch_2}. Although techniques such as sharding and off-chain storage improve transaction throughput, their implications for query processing and data accessibility across interoperable blockchains remain largely unexplored. Future research should examine how scalable architectures can support efficient indexing, cross-chain queries, and consistent data retrieval in multi-chain blockchain ecosystems \cite{query_ch_1}.
% Multiple blockchains may contain diverse data, and other nodes may need to access and share this information. To query data, the entities also need to know what kind of data are available and what data they can query. This points to the need to create an indexing mechanism to guide the systems in searching for data in the correct locations/nodes.

\subsection{Privacy Concerns}
% Why are there public and private blockchains 
% What happens to the private data communicated between parachains 
% How good is storing the data on the relay chain, which everyone has access 

Blockchains were primarily public in the initial years of their development. Anyone can join and leave the network at any time. All network members can see all transactions in the network, and everyone has equal rights. With the increasing digitisation of financial, commercial, and healthcare services, privacy preservation has become a critical requirement for modern digital systems. In blockchain-based environments, several privacy challenges remain unresolved, including user identity exposure, transaction linkability, key management, and regulatory compliance, which must be addressed to enable secure and trustworthy interoperable systems \cite{privacy_ch_1,privacy_ch_2}. As blockchains have become popular, organisations have ventured into blockchain technology and do not want their data to be publicly available, which has led to the growth of private blockchains. Private blockchains restrict access to only approved members with access controls. In both cases, the underlying blockchain was the same. The situation is more complex with respect to interoperable blockchains and even more so with multiple heterogeneous blockchains \cite{pace}. Consider the case study discussed in the paper. All nodes on the relay chain are aware of the communication/transaction between two parachains, primarily with the gossip process involved in XCMP. Moreover, the validators of the receiver parachain push the message to the relay chain when they validate the message. All parachains have access to the relay chain and all the information on the relay chain. Although the relay chain provides security for interchain communication, having all parachain communications on the relay chain may violate the privacy of the participating entities.
\subsection{Security and Governance Concerns}
% o	Share information and assets
% o	What happen if there are vulnerabilities in the chain
% o	If here is vulnerability in one chain they may be able to exploit the other
% o	Governnace and decision making
% o	What if the validators or miners are corrupt

Interoperable systems communicate with other systems by exchanging assets or information. The vulnerabilities in one system may affect others due to their connections. If a blockchain has vulnerabilities and is part of an interoperable environment, it may create insecurity for all other connected blockchains. Malicious entities in the system may exploit other entities in the chain, leading to inconsistencies in assets or manipulation of information in the mainnet \cite{security_ch_2}. Recent generations of blockchain systems do not use Bitcoin's proof-of-work (PoW) but use consensus protocols managed by fewer entities. Although there are mechanisms to secure the system, any vulnerability in the validation process may affect the entire architecture of the multilayer blockchain. Establishing accountability and clear governance structures across interconnected chains remains a significant challenge, particularly when networks operate under different consensus rules and organisational structures \cite{gove_ch1}. Cross-chain protocols must include mechanisms for dispute resolution and trust establishment, such as time-locked contracts and oracle-based verification for cross-chain transactions. Systems must also mitigate the risks posed by different consensus protocols and validation mechanisms, which can introduce inconsistencies when integrating multiple blockchains \cite{security_ch_3}. Another important security concern arises from cross-chain bridges, which are widely used to enable asset transfers between independent blockchain networks. Several high-profile security incidents in recent years have demonstrated that vulnerabilities in bridge protocols or their validation mechanisms can lead to significant asset losses, highlighting the need for robust cryptographic verification and secure cross-chain communication protocols \cite{security_ch4}. Furthermore, security vulnerabilities in smart contracts and the absence of clear regulatory governance mechanisms remain significant barriers to the large-scale deployment of interoperable blockchain systems \cite{security_ch_1}.

% Consider Polkadot parachains which are independent entities with their own governance and validation process. If incorrect or corrupt information flows from one parachain to the relay chain, it may compromise the system's integrity.
\subsection{Scalability}
% o	Limit on the number of systems they can connect
% o	In the case of Polkadot, the maximum number of parachains is 100
% o	Avalanche can handle an infinite number
Blockchain technology is growing, and every year, new blockchain platforms are introduced with new features. There are hundreds, if not thousands, of blockchain platforms available and no accurate details about the number of platforms in the market. Since no blockchain can fully satisfy the needs of every application, new platforms may enter the market with some new features \cite{scalability_ch_2}. The layer-0 blockchains support scalability; the question is how many different chains they can support simultaneously \cite{scalability_ch_3}. Polkadot currently supports a limited number of parachains (approximately one hundred), which is the maximum number, whereas Avalanche can support an infinite number of sidechains. Scalability becomes more complex when considering the different consensus protocols of the blockchains and their varying transaction processing rates. The limitations of one system should not affect the performance of other systems while permitting scalability. As the number of participating nodes and transactions in a blockchain network increases, challenges related to storage capacity, data processing speed, and transaction latency become more pronounced, directly affecting the system scalability. Consequently, improving scalability has become a major research focus in blockchain systems, with proposed solutions aimed at enhancing throughput, reducing operational costs, and improving overall network efficiency \cite{scalability_ch_1}.

\subsection{Standardisation of Interoperability Protocols}
% many protocols and no common standard
% Fragmentation of development
% need to have something like OSI layers in networks.
Interoperability remains a significant challenge for blockchain ecosystems, with the absence of universally accepted standards posing major hurdles \cite{std_1}. Current solutions such as XCMP in Polkadot and IBC in Cosmos are designed to address interoperability within their ecosystems but are platform-specific, limiting their application across heterogeneous blockchains. Despite the rapid growth of blockchain technologies, many solutions have been developed in isolation without common protocols or cryptographic standards, resulting in fragmented ecosystems that limit interoperability between platforms. The absence of widely accepted blockchain-specific standards has slowed the practical deployment of interoperability solutions, with most existing approaches still focusing primarily on cryptocurrency-based applications rather than broader data-driven use cases \cite{std_2}. This lack of standardisation leads to fragmentation, as each blockchain implements its own protocols, messaging formats, and data structures. The development of industry-wide standards could help address this issue by enabling seamless integration and communication across blockchains. Moreover, standardised protocols would reduce redundancy in development and allow blockchain platforms to focus on innovation rather than reinventing interoperability mechanisms \cite{std_3}.

The aforementioned challenges are closely interconnected and collectively influence the development of interoperable blockchain ecosystems. For example, effective data management and indexing mechanisms are essential for enabling cross-chain querying, whereas privacy and security considerations must be carefully balanced with the interoperability requirements. Similarly, governance mechanisms and standardised interoperability protocols are necessary to coordinate interactions among independent blockchain networks. Addressing these challenges in an integrated manner is essential for enabling scalable, secure, and trustworthy interoperable blockchain systems in future applications.

\section{Conclusion} \label{sec: conclusion}
This study has shown that interoperable blockchains offer a practical route beyond the limitations of isolated single-layer blockchain systems by enabling data, asset, and state exchange across heterogeneous networks. Our platform-oriented analysis highlights that existing ecosystems already provide a range of interoperability mechanisms, including relay-chain, hub-based, bridge-based, and subnet-based approaches, each with different strengths and trade-offs. For practitioners and system designers, the main implication is that interoperability should not be treated as a purely technical add-on: platform choice directly affects security assumptions, scalability, governance, privacy, and integration complexity. The conceptual supply-chain case study illustrates how such architectures can coordinate multi-entity interactions across heterogeneous blockchain environments, but it also has clear limitations, as it is intended only to demonstrate architectural feasibility and does not provide prototype validation or performance benchmarking. Overall, interoperable blockchain systems present significant potential, but real deployment requires careful attention to cross-chain security, data management, privacy protection, governance, and protocol standardisation. Future research should therefore focus on secure and verifiable cross-chain communication, scalable indexing and query mechanisms for multi-chain environments, governance and dispute-resolution models across independent networks, and common interoperability standards that can support broader real-world adoption.

\noindent {\bf Competing Interests:} The authors declare no conflicts of interest.
\newline
\newline
\noindent {\bf Funding Information:} Supported in part by the Engineering and Physical Sciences Research Council ``Digital Economy" program: EP/V042521/1 and EP/V042017/1
\newline
\newline
\noindent {\bf Author Contribution:} methodology, S.W. and K.A.-D.; validation, S.W., K.A.-D., E.S and O.R; formal analysis, S.W., and K.A.-D.; investigation, S.W., K.A.-D., and E.S; draft, S.W., K.A.-D., and E.S; writing---review and editing, Y.L., E.S., O.R., and R.R; visualisation, S.W. and K.A.-D.; supervision, Y.L., E.S., R.R., and O.R.; project administration, R.R. and O.R.; funding acquisition, R.R., and O.R. All authors have read and agreed to the published version of the manuscript.
\newline
\newline
\noindent {\bf Data Availability Statement:} Not Applicable.
\newline
\newline
\noindent {\bf Research Involving Human and /or Animals:} Not Applicable.
\newline
\newline
\noindent {\bf Informed Consent Statement:} Not Applicable.

\bibliographystyle{ieeetr}
\bibliography{survery_paper}

\end{document}